\documentclass[twocolumn,9pt]{article}

\usepackage[utf8]{inputenc}
\usepackage{siunitx,color,xfrac,authblk,titlesec}
\usepackage[font=small]{caption}
\usepackage[version=4]{mhchem}
\usepackage[compress,superscript,biblabel]{cite}
\usepackage{tabularx}
\newcolumntype{C}{>{\centering\arraybackslash}X}

\graphicspath{{Figures/}}
\renewcommand{\thefigure}{\textbf{\arabic{figure}}}

\renewcommand{\thetable}{\textbf{\arabic{table}}}
\titleformat*{\section}{\large\bfseries}

\makeatletter
\renewenvironment{abstract}{%
    \if@twocolumn
      \section*{\abstractname}%
    \else 
      \begin{center}%
        {\bfseries \normalsize\abstractname\vspace{\z@}}
      \end{center}%
      \quotation
     \fi}
    {\if@twocolumn\else\endquotation\fi}
\makeatother

\begin{document}
 \title{\vspace{-5em}\textbf{\Large Origin of metallic conductance in a single-component molecular organic crystal}}
 \author{Tobias Schlöder}
 \author{Wolfgang Wenzel\thanks{q}}
 \affil{\textit{Karlsruhe Institute of Technology, Institute of Nanotechnology, Hermann-von-Helmholtz-Platz 1, 76344 Eggenstein-Leopoldshafen}}
 \date{}
  
 \twocolumn[
  \begin{@twocolumnfalse}
   \maketitle
   \vspace{-1em}
   \begin{abstract}
    Since the discovery of the TTF-TCNQ charge transfer complex as the first metallic material composed of molecules, many other molecular metals were reported. It was however only recently that the first metal-free single-component organic metal was characterized by Kobayashi et al. Although the measured properties of a poorly crystalline sample clearly showed metallic behavior, the crystal structure itself could not be solved, so that the conduction mechanism in this material is still unknown. Here, we present the results of theoretical crystal structure prediction calculations for the TED molecule, accompanied by electronic DOS and band structure calculations which indicate band transport.
    \\
   \end{abstract}
  \end{@twocolumnfalse}
  ]
 {
  \renewcommand{\thefootnote}%
   {\fnsymbol{footnote}}
  \footnotetext[1]{wolfgang.wenzel@kit.edu}
 } 

 \section*{Introduction}
 Metallic behavior is classically associated with inorganic elements and compounds, and metals have since long been used as materials for tools due to properties such as  malleability, ductility, toughness as well as resistance towards oxidation or other kinds of degradation. From a physics viewpoint, the modern definition of a metallic solid (or cluster) is based on its electronic structure and requires the absence of a band gap, i.e. a finite electronic density of states (eDOS) at the Fermi level. This distinguishes metals from other solids for which the eDOS vanishes at the Fermi level, such as insulators (large gap between the valence and conductance band) or semiconductors (small band gap with respect to thermal energy). Organic molecules usually display large HOMO-LUMO gaps, which typically results in insulating behavior in the solid state. One class of  known exceptions to this rule are  organic solids comprising 1:1 charge-transfer (CT) complexes, which were first discovered in 1973 at the example of tetrathiafulvalene (TTF) and tetracyanoquinodimethane.\cite{Ferraris1973} These CT based organic metals inherently consist of two molecules, a donor as well as an acceptor, and all other known single-component molecular metals contain at least one metallic element such as e.g. the nickel complexes \ce{[Ni(ptdt)2]}\cite{Kobayashi1999} and \ce{[Ni(tmdt)2]}.\cite{Tanaka2001}  Neutral radical molecules have also since long been discussed as possible candidates\cite{Haddon1975} but the usually poor orbital overlap in the corresponding crystals leads to narrow partially filled bands of localized states.
 
 It was only in 2017 when metallic behavior was found for the first time for a single component pure organic, i.e. metal-free, material consisting of only one type of molecule by Kobayashi et. al., for which 'copper-like' conductivities of \SI{1000}{S.cm^{-1}} at \SI{50}{K} and \SI{530}{S.cm^{-1}} at \SI{300}{K} were measured.\cite{Kobayashi2016,Kobayashi2018} This  tetrathiafulvalene extended dicarboxylate radical (TED, see fig.~\ref{fig:mol}), is comprised of a neutral zwitterion containing a radical cationic bis-tetrathiafulvalene unit, whose positive charge is compensated by a carboxylate group stabilized via an intramolecular hydrogen bond. SEM images revealed a flat layered structure of the material, but no detailed information about the solid-state structure is available as attempts to obtain single crystals of the TED radical have failed to date. Hence, the conduction mechanism of this new class of organic metals is presently unknown.
  
 \section*{Results}
\begin{figure}[t!]
  \includegraphics[width=\linewidth]{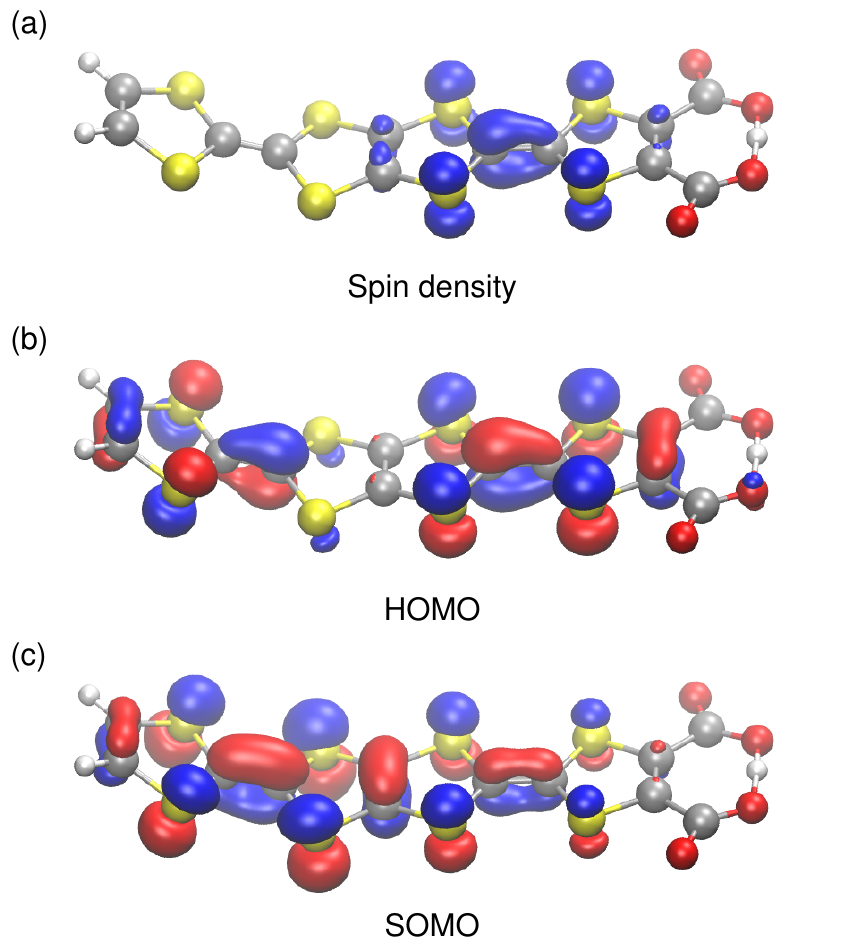}
  \caption{DFT-optimized molecular structure of TED together with isosurfaces for the (a) Spin density, (b) HOMO, and (c) SOMO. Color code: carbon (gray), sulfur (yellow), oxygen (red), hyrdogen (white).}
  \label{fig:mol}
 \end{figure}

 In order to be able to compute the electronic structure of the material, we have, as a first step, computed the structure of the TED radical in the gas phase using molecular quantum-chemical calculations at the DFT level, and our results are in good agreement with those previously reported.\cite{Kobayashi2016} The two condensed TTF units are slightly bent with respect to each other in the gas phase and the radical cationic part of the zwitterion is localized on the TTF unit bearing the carboxyl groups as shown by a plot of the spin density (fig.~\ref{fig:mol}a). When we simulated the molecule in a fictitious solid environment with permittivities of $\epsilon \ge 12$ using the COSMO model, structure optimization yielded almost perfectly planar structures, in which the positive charge is delocalized over both TTF units (see also table \ref{tab:mol_struct} in the appendix). The corresponding increase of the dipole moment explains why the molecule is bent in vacuum. An analysis of the electronic structure of the molecule showed that the two highest (partially) occupied molecular orbitals (HOMO and SOMO, see fig. \ref{fig:mol}b and c) have $\pi$-character and their partial occupation makes them ideally suited for $\pi$-stacking.

 \begin{figure*}
  \includegraphics[width=\linewidth]{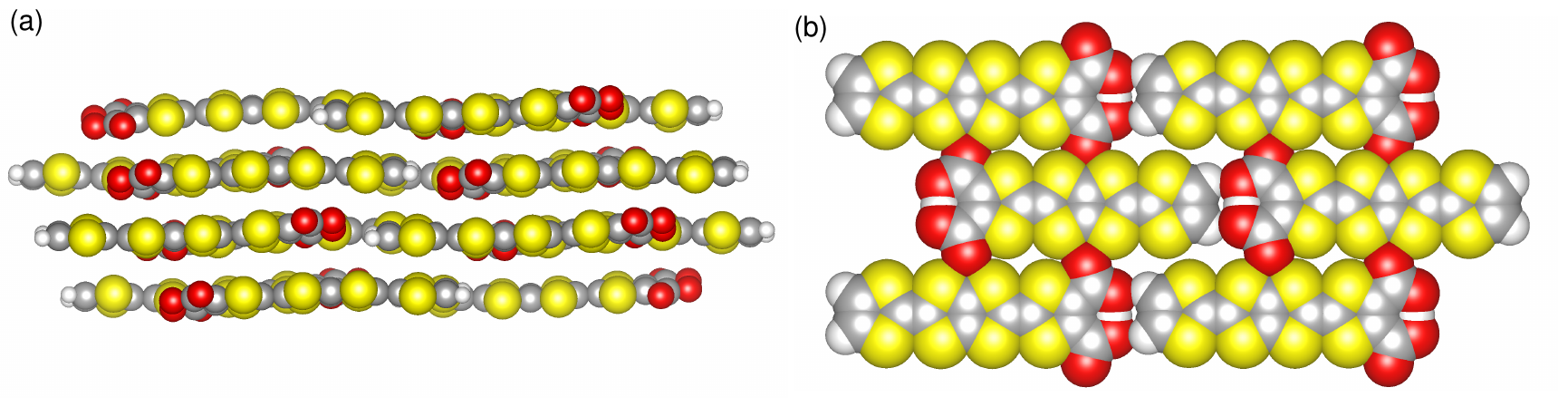}
  \caption{Two views of the predicted crystal structure of TED: (a) sideview [(010) projection] showing the layered structure; (b) topview showing the arrangement within the layers. Color code: carbon (gray), sulfur (yellow), oxygen (red), hyrdogen (white).}
  \label{fig:cryst}
 \end{figure*}
 
 We then performed a crystal structure prediction based on this structure using the USPEX code by Oganov et al.\cite{Oganov2006,Oganov2011,Lyakhov2013,Zhu2012} with $Z=4$ molecules per unit cell, which is the minimum necessary for an electrostatically favorable 3D periodic arrangement of dipoles. The energy of all structure models was calculated using periodic plane-wave DFT calculations with each molecule bearing one unpaired electron. The resulting best structure is shown in figure~\ref{fig:cryst} and was found to belong to the monoclinic space group $P2_1/c$ (no. 14; optimized lattice parameters: $a$~=~\SI{7.217}{\angstrom}, $b$~=~\SI{12.989}{\angstrom}, $c$~=~\SI{16.561}{\angstrom} and $\beta$~=~\ang{89.67}; see table \ref{tab:cryst_para}). All atoms are located on the general Wyckoff position 4e so that all four molecules in the unit cell are symmetry-equivalent. The structure can be described as an arrangement of $\pi$-stacked layers in approximatively [101] direction where each unit cell contains two of the layers, resulting in a calculated interlayer distance of $d_{calc}$~=~\SI{3.315}{\angstrom}. This value agrees well with the most intense reflection in the calculated XRD powder diffractogram (at $2\theta$~=~\ang{26.87} for Cu\textsubscript{$\alpha$} irradiation; see figure \ref{fig:xrd}) and is furthermore in excellent agreement with the main feature of published measured XRD pattern at $d_{exp}$~=~\SI{3.32}{\angstrom}. Within the layers, the molecules align to optimize the electrostatic interaction between the partial charges of the single molecules: where within one strand of molecules the dipoles all point in the same direction, their direction are inverted in the neighboring strands; in addition there are further close contacts between the carbonyl oxygen atoms and the central sulfur atoms of the positively charged bis-TTF unit.
 
 \begin{figure*}[t]
  \includegraphics[width=\linewidth]{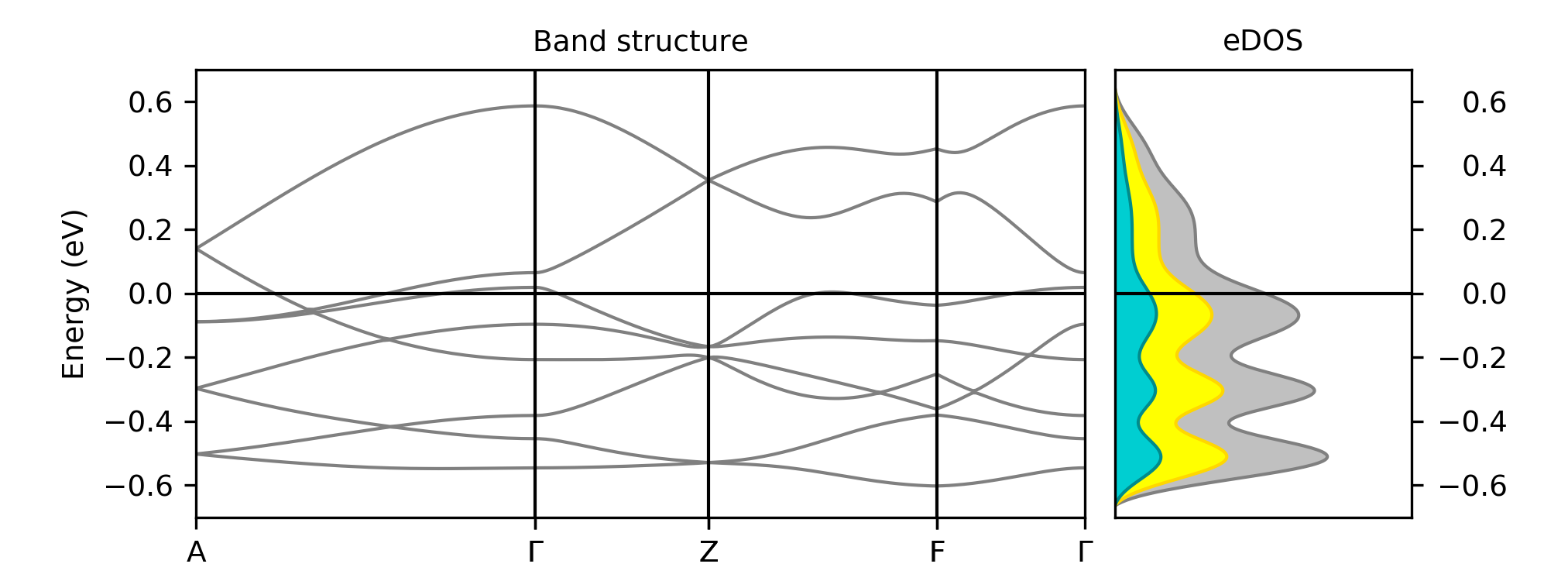}
  \caption{Electronic densitiy of states of crystalline TED andtable band structure on the interlayer ($\Gamma$--A) as well as on an intralayer k-path ($\Gamma$--Z--F--$\Gamma$) in the vicinity of the Fermi level ($E_F = 0$); colors for partial eDOS: C (turquoise), S (yellow), others (gray). Only $\alpha$ electrons are shown.}
  \label{fig:dos}
 \end{figure*}
 In the next step, the electronic structure of this phase was analysed by eDOS and band structure calculations. As shown in the right panel of fig.~\ref{fig:dos}, a finite eDOS was found for the $\alpha$-electrons at the Fermi level, where the main contribution to corresponding bands comes from the 2p~(C) and 2p~(S) orbitals. No substantial contribution of other orbitals to these bands was observed, and the remaining eDOS is due to delocalized states between the molecules. These findings are furthermore in good agreement with the electronic structure of the isolated molecules whose frontier orbitals are of $\pi$(C--S)/$\pi$(C--C) character. Together with the small intralayer distance of \SI{3.32}{\angstrom}, it is reasonable to assume a good overlap between the $\pi$-systems of the individual molecules which would lead to a high conductivity perpendicular to the layers, i.e. in [101] direction.
 
 This question was further investigated by calculating the electronic band structure along k-lines which were chosen to represent paths between [$\Gamma$--A(\textonehalf,0,\textonehalf)] as well as within in the layers [$\Gamma$--Z(0,\textonehalf,0)--F(0.09495,0,--\textonehalf)--$\Gamma$], where the k-points of the latter line describe a triangle with in the (010) layers (left panel of figure~\ref{fig:dos}). Corroborating the initial assumption, a steep band crosses the Fermi level on the interlayer path, which results in a high conductivity in perpendicular to the layers. By contrast, the much smaller slopes of the bands at the Fermi level, hint at low --- but nevertheless nonvanishing --- intralayer conductivity. 

 A more detailed estimation of the conductivity of the crystals was obtained from $\sigma=e^2\tau/m^{*}$, where $e$ is the elementary charge, $\tau$ the relaxation time and $m^{*}$ the effective mass of the electron. According to \[ m^{*}=\hbar^2\left(\frac{\partial^2E}{\partial k^2}\right)^{-1},\] an effective mass of $m^{*}$~$\approx$~\SI{1.7}{m_{el}} was calculated from the second derivative of the band crossing the Fermi level on the $\Gamma$--A path. Together with a typical relaxation time in organic single crystals of \SI{200}{\femto\second} at \SI{300}{\kelvin},\cite{Ishii2013} an upper limit of $\sigma$~$\approx$~\SI{2e6}{S.cm^{-1}} was estimated for the conductivity of single crystal TED. Additional scattering at the grain boundaries of the poorly crystalline sample for which the conductivity measurements were made, account for the much lower value found in the experiment.
 
 \section*{Summary}
 In conclusion, our predicted crystal structure is consistent with the experimental findings: Both the experimentally observed 'flat layered structure' and the main feature of the powder diffractogram agree well   with our calculations, and our structure data could be used for solving the crystal structure of TED by Rietveld methods. In addition, our electronic structure calculations support the description of the TED radical as the first example of an organic metal consisting of only one molecules. The measured copper-like conductivities can thus unambiguously assigned to a true metallic behavior of TED. Based on these results, it can be expected that similarly high conductivities can be achieved for other $\pi$-stacked organic crystals with an open-shell $\pi$-system which might thus serve as a design principle when searching for highly conducting pure organic materials.
 
 \section*{Computational details}
 All molecular DFT calculations were carried out within the generalized gradient approximation (GGA) using the B3LYP hybrid functional\cite{Becke1988,Becke1993,Lee1988} together with the RI approximation\cite{Eichkorn1995} as implemented in the Turbomole 7.3. suite of programs.\cite{tm73} The def2-TZVP valence basis sets and the corresponding auxiliary basis sets were used for all elements.\cite{Weigend2005,Eichkorn1997} The effect of a dielectric medium was simulated using the conductor-like screening model (COSMO)\cite{Klamt1993} with different permittivities. The prediction of the crystals structure was done using the USPEX code,\cite{Oganov2006,Oganov2011,Lyakhov2013,Zhu2012} based on an evolutionary algorithm which features local optimizations of a large number of trial structures. Starting from an initial population of 30 randomly generated structures ('individuals') in the first iteration ('generation'), each subsequent generation with 25 individuals was created by applying different variation operators to the best structures of the previous generation. This was repeated until convergence of the associated energies which were evaluated using an external code:  Six successive structure optimizations with increasing accuracy were performed for each individual generated by USPEX, using version 5.4.1 of the Vienna ab initio structure package (VASP).\cite{Kresse1993,Kresse1994,Kresse1996,Kresse1996a} The exchange correlation energies were calculated at the spin-polarized DFT level using the GGA functional PBE,\cite{Perdew1992,Perdew1997} and the obtained energies were corrected for long-range dispersive interactions using Grimme’s D3 method with a Becke-Johnson damping.\cite{Grimme2010,Grimme2011} Plane wave basis were used to construct the Kohn-Sham orbitals, the core regions of which were described by projector-augmented wave (PAW) potentials. A Gaussian smearing of $\sigma$~=~0.1 and $\Gamma$-centered k-grids were used in all calculations. For the sixth and final structure optimizations, a k-resolution of 0.05~2$\pi$~\SI{}{\angstrom^{-1}} was used together with an energy cutoff of \SI{520}{eV}, and energies were converged to \SI{5e-5}{eV}. The last structure relaxation was then followed by a single point energy calculation using a k-resolution of 0.02~2$\pi$~\SI{}{\angstrom^{-1}}, an energy cutoff of 600 eV and a convergence criterion of \SI{2e-5}{eV}. The predicted best structure was then optimized within the restrictions of the $P2_1/c$ space group, and the resulting atomic positions were used for subsequent eDOS and band structure calculations. The eDOS was computed using an energy cutoff of \SI{600}{eV} and a k-resolution of 0.0125~2$\pi$~\SI{}{\angstrom^{-1}}. The self-consistent charge densities of the eDOS calculation were then used to obtain the eigenvalues at k-points used in the band structure plot.
 
 \section*{Acknowledgments}
 The authors would like to thank Dr. Artem Fediai for support with calculating conductivities. All calculations in this work were performed on the supercomputer ForHLR funded by the Ministry of Science, Research and the Arts Baden-Württemberg and by the Federal Ministry of Education and Research. The authors would further like to acknowledge funding by the 'Virtual materials design' (VirtMat) initiative at KIT.
   
 \bibliographystyle{h-physrev.bst}

 \newpage
 \section*{Appendix}
 \setcounter{table}{0}
 \renewcommand{\thetable}{\textbf{A\arabic{table}}}
 \setcounter{figure}{0}
 \renewcommand{\thefigure}{\textbf{A\arabic{figure}}}

 \begin{table}[h]
  \caption{Calculated molecular structures of TED at the B3LYP/def2-TZVP level of theory (coordinates in \SI{}{\angstrom})}
  \label{tab:mol_struct}
  \small
  \begin{tabularx}{\linewidth}{CCCCCCC} 
   & \multicolumn{3}{c}{In vacuo} & \multicolumn{3}{c}{COSMO model\textsuperscript{a}}\\
   \multicolumn{1}{c}{Atom} & x & y & z & x & y & z\\\hline
   C1 & 2.148 & -0.173 & 4.870 & 2.174 & 0.087 & 4.894\\
   C2 & 2.542 & 0.704 & 7.275 & 2.584 & 0.235 & 7.448\\
   C3 & 3.757 & 0.625 & 6.736 & 3.801 & 0.160 & 6.908\\
   C4 & 1.599 & -0.491 & 3.679 & 1.620 & 0.040 & 3.651\\
   C5 & 1.218 & -0.515 & 1.154 & 1.236 & -0.052 & 1.111\\
   C6 & -0.017 & -0.436 & 1.702 & -0.008 & 0.025 & 1.663\\
   C7 & -0.387 & -0.256 & -0.831 & -0.390 & -0.055 & -0.874\\
   C8 & -0.942 & -0.158 & -2.100 & -0.944 & -0.072 & -2.124\\
   C9 & -1.312 & -0.024 & -4.639 & -1.335 & -0.129 & -4.687\\
   C10 & -2.547 & 0.051 & -4.094 & -2.570 & -0.051 & -4.141\\
   C11 & -3.914 & 0.186 & -4.733 & -3.929 & 0.001 & -4.778\\
   C12 & -0.833 & -0.002 & -6.099 & -0.886 & -0.194 & -6.138\\
   O1 & -4.862 & 0.232 & -3.969 & -4.903 & 0.071 & -4.042\\
   O2 & -3.969 & 0.241 & -6.021 & -4.008 & -0.030 & -6.074\\
   O3 & -1.726 & 0.101 & -6.983 & -1.780 & -0.171 & -7.039\\
   O4 & 0.388 & -0.091 & -6.210 & 0.335 & -0.264 & -6.312\\
   S1 & 3.888 & -0.039 & 5.124 & 3.905 & 0.039 & 5.173\\
   S2 & 1.192 & 0.134 & 6.320 & 1.221 & 0.204 & 6.362\\
   S3 & 2.566 & -0.820 & 2.227 & 2.593 & -0.077 & 2.186\\
   S4 & -0.152 & -0.646 & 3.433 & -0.121 & 0.091 & 3.389\\
   S5 & -1.354 & -0.218 & 0.622 & -1.359 & 0.043 & 0.586\\
   S6 & 1.334 & -0.388 & -0.570 & 1.345 & -0.124 & -0.611\\
   S7 & 0.023 & -0.173 & -3.541 & 0.014 & -0.166 & -3.581\\
   S8 & -2.646 & -0.010 & -2.355 & -2.666 & 0.003 & -2.392\\
   H1 & 2.325 & 1.068 & 8.268 & 2.371 & 0.312 & 8.504\\
   H2 & 4.674 & 0.917 & 7.226 & 4.731 & 0.167 & 7.459\\
   H3 & -2.990 & 0.181 & -6.492 & -3.032 & -0.093 & -6.542\\
   \multicolumn{7}{l}{\textsuperscript{a}{Using $\epsilon$~=~12}}
  \end{tabularx}\normalsize
 \end{table}
 
 \newpage
 \begin{table}[ht]
  \caption{Structure parameters of the predicted crystal structure of TED at the PBE-D3/pw(PAW~P) level of theory}
  \label{tab:cryst_para}
  \small
  \begin{tabular}{ll}
   Crystal system & monoclinic \\
   Space group & $P2_1/c$ -- No. 14 \\
   Lattice constants & $a$~=~\SI{7.217}{\angstrom}\\
   & $b$~=~\SI{12.989}{\angstrom}\\
   & $c$~=~\SI{16.561}{\angstrom}\\
   & $\beta$~=~\ang{89.67} \\
   Cell volume & $V$~=~\SI{1552.45}{\angstrom^3} \\
   Formula units & $Z$~=~4 \\
   Density & $\varrho$~=~\SI{2.001}{g.cm^{-3}} \\
  \end{tabular}
  
  \begin{tabularx}{\columnwidth}{CCCCCCCCCC}
   \\ \multicolumn{10}{c}{Atomic positions (fractional coordinates)}\\ 
   \multicolumn{1}{c}{Atom} & Site & x & y & z & \multicolumn{1}{c}{Atom} & Site & x & y & z \\\hline
   C1 & 4e & 0.616 & 0.278 & 0.228 & O3 & 4e & 0.509 & 0.497 & 0.290\\
   C2 & 4e & 0.646 & 0.351 & 0.160 & O4 & 4e & 0.547 & 0.622 & 0.197\\
   C3 & 4e & 0.626 & 0.456 & 0.157 & S1 & 4e & 0.721 & 0.290 & 0.072\\
   C4 & 4e & 0.556 & 0.530 & 0.220 & S2 & 4e & 0.684 & 0.516 & 0.067\\
   C5 & 4e & 0.746 & 0.403 & 0.016 & S3 & 4e & 0.134 & 0.711 & 0.112\\
   C6 & 4e & 0.190 & 0.597 & 0.062 & S4 & 4e & 0.164 & 0.483 & 0.117\\
   C7 & 4e & 0.067 & 0.650 & 0.200 & S5 & 4e & 0.017 & 0.288 & 0.715\\
   C8 & 4e & 0.078 & 0.544 & 0.202 & S6 & 4e & 0.006 & 0.483 & 0.289\\
   C9 & 4e & 0.060 & 0.403 & 0.663 & S7 & 4e & 0.220 & 0.516 & 0.541\\
   C10 & 4e & 0.144 & 0.404 & 0.588 & S8 & 4e & 0.185 & 0.291 & 0.533\\
   C11 & 4e & 0.291 & 0.457 & 0.453 & H1 & 4e & 0.538 & 0.397 & 0.296\\
   C12 & 4e & 0.277 & 0.353 & 0.449 & H2 & 4e & 0.347 & 0.504 & 0.404\\
   O1 & 4e & 0.564 & 0.312 & 0.298 & H3 & 4e & 0.318 & 0.306 & 0.398\\
   O2 & 4e & 0.646 & 0.185 & 0.214\\
  \end{tabularx}\normalsize
 \end{table}

 \begin{figure*}[t!]
  \includegraphics[width=\linewidth]{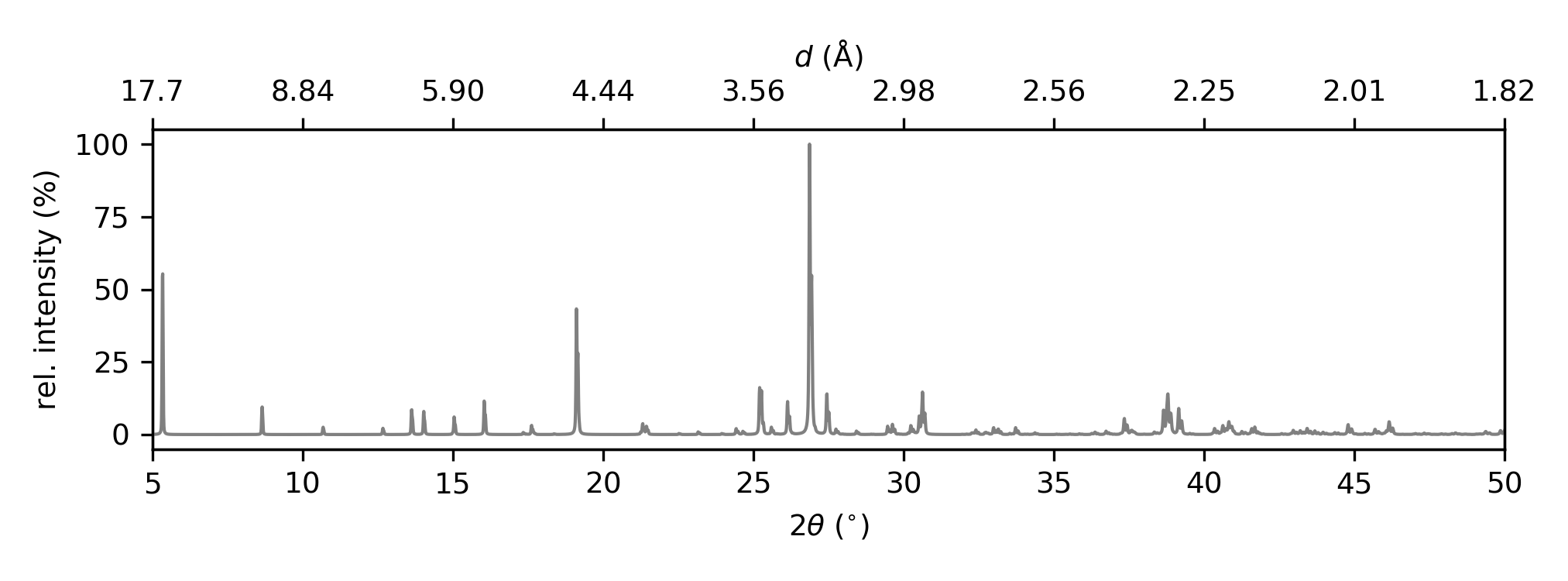}
  \caption{Calculated XRD powder diffractogram of crystalline TED for Cu\textsubscript{$\alpha$} irradiation.}
  \label{fig:xrd}
 \end{figure*}
 
\end{document}